\documentstyle[twoside,fleqn,psfig,babel,english,hyperon99_paper]{article}

\newcommand{\AmS}{{\protect\the\textfont2
  A\kern-.1667em\lower.5ex\hbox{M}\kern-.125emS}}

\title{Particle - antiparticle asymmetries in the production of 
baryons in 500 GeV/$c$ $\pi^-$-nucleon interactions}

\author{J.C. Anjos \address{Centro Brasileiro de Pesquisas F\'{\i}sicas,\\
        Rua Dr. Xavier Sigaud 150 - Urca\\
        22290-180 Rio de Janeiro, Brazil}\\ 
        Representing the Fermilab E791 Collaboration}
       
\begin{document}

\begin{abstract}
We present the Fermilab E791 measurement of baryon - antibaryon 
asymmetries in the production of $\Lambda^{\circ}$, $\Xi$, $\Omega$ and 
$\Lambda_c$ in 500 GeV/$c$ $\pi^-$-nucleon. Asymmetries have been 
measured as a function of $x_F$ and $p_T^2$ over the range 
$-0.12 < x_F < 0.12$ and $p_T^2 < 4$ (GeV/$c$)$^2$ for hyperons and 
$-0.1 < x_F < 0.3$ and $p_T^2<8$ (GeV/$c$)$^2$ for the $\Lambda_c$ 
baryons. We observe clear evidence of leading particle effects 
and a basic asymmetry even at $x_F=0$. These are the first high statistics 
measurements of the asymmetry in both the target and beam fragmentation 
regions in a fixed target experiment.

\end{abstract}

\maketitle


Particle - antiparticle asymmetry is an excess in the production rate 
of a particle over its antiparticle (or vice-versa). It can be quantified 
by means of the asymmetry parameter
\begin{equation}
A = \frac{N - \bar{N}}{N + \bar{N}} \; ,
\label{eq1}
\end{equation}
where $N$ ($\bar{N}$) is the number of produced particles (antiparticles).

Measurements of this parameter show leading particle effects, 
which are manifest as an enhancement in the production rate of particles 
which have one or more valence quarks in common with the initial (colliding) 
hadrons, compared to that of their antiparticles which have fewer valence quarks 
in common.

Other effects, such as associated production of meson and baryons can also 
contribute to a non-zero value of the asymmetry parameter.

Leading particle effects in charm hadron production have been extensively 
studied in recent years from both the experimental~\cite{experiment} and 
theoretical point of view~\cite{theory}. The same type of leading particle effects 
are expected to appear in strange hadron production. Although previous 
reports of global asymmetries in $\Lambda^{\circ}$, $\Xi$ and $\Omega$ 
hadroproduction already exist~\cite{hyperon}, there is a lack of a systematic 
study of light hadron production asymmetries. 

From a theoretical point of view, models which can account for the presence 
of leading particle effects in charm hadron production use some kind of 
non-perturbative mechanism for hadronization, in addition to the perturbative 
production of charm quarks~\cite{theory}.

Given E791's $\pi^-$ beam incident on nucleon targets, strong differences 
are expected in the asymmetry in both the $x_F<0$ and $x_F>0$ regions. 
In particular, as $\Lambda$ (or $\Lambda_c$) baryons are double leading 
in the $x_F<0$ region while both $\Lambda$ ($\Lambda_c$) and 
$\bar{\Lambda}$ ($\bar{\Lambda_c}$) are leading in the $x_F>0$ region, a 
growing asymmetry with $\left|x_F\right|$ is expected in the negative 
$x_F$ region and no asymmetry is expected in the positive $x_F$ region.
$\Xi^-$ baryons are leading in both the positive and negative $x_F$ 
regions, whereas $\Xi^+$ are not. Thus a growing asymmetry with $\left| 
x_F\right|$ is expected in this case. $\Omega^{\pm}$ are both non 
leading, so no asymmetry is expected at all.


Experiment E791 recorded data from 500 GeV/c $\pi^-$ 
interactions in five thin foils (one platinum and four diamond) 
separated by gaps of 1.34 to 1.39 cm. Each foil was approximately 0.4\% of 
a pion interaction length thick (0.6 mm for the platinum foil and 1.5 mm 
for the carbon foils). A complete description of the E791 spectrometer 
can be found in Ref.~\cite{e791}.

An important element of the experiment was its extremely 
fast data acquisition system \cite{e791} which, combined with a 
very open trigger requiring a beam particle and
a minimum transverse energy deposited in the calorimeters, was used
to record a data sample of $2 \times 10^{10}$ interactions.

The E791 experiment reconstructed more than $2 \times 10^5$ charm 
events and many millions of strange baryons.


Hyperons produced in the carbon targets and with decay point 
downstream the SMD planes were kept for further 
analysis. 

$\Lambda^{\circ}$ were selected in the $p\pi^-$ and c.c. decay 
mode. All combinations of two tracks with an {\it a priori} Cherenkov 
probability of being identified as a $p\pi^-$ combination were selected for 
further analysis if the tracks have a distance of closest approach less than 
$0.7$ cm from the decay vertex. In addition, the invariant mass was required 
to be between $1.101$ and $1.127$ GeV/$c^2$, the ratio of the momentum of the proton 
to that of the pion was required to be larger than 2.5 and the reconstructed 
$\Lambda^{\circ}$ decay vertex must be downstream of the last target. The 
impact parameter must be less than $0.3$ cm if the particle decays within the 
first $20$ cm and 0.4 if decaying more than $20$ cm downstream of the target region.

$\Xi$'s were selected in the $\Lambda^{\circ}\pi^-$ and c.c. decay mode 
and at the same time $\Omega$'s were selected in the $\Lambda^{\circ}K^-$ and 
c.c. channel. Starting with a $\Lambda^{\circ}$ candidate, a third distinct track 
was added as a possible pion or kaon daugther. All three tracks were required 
to be only in the drift chamber region. Cuts for the daughter 
$\Lambda^{\circ}$ were the 
same as above, except for that on the impact parameter. The invariant mass 
for the three track combination was required to be between $1.290$ and 
$1.350$ GeV/$c^2$ and $1.642$ and $1.702$ GeV/$c^2$ for $\Xi$ and $\Omega$ 
candidates respectively. In addition, the $\Xi$ and $\Omega$ decay 
vertices were required to be upstream the $\Lambda^{\circ}$ decay vertex and 
downstream the SMD region. For $\Omega$'s, the third track had to have a 
clear kaon signature in the Cherenkov counter.

\begin{figure}[hctb]
\centerline{\psfig{figure=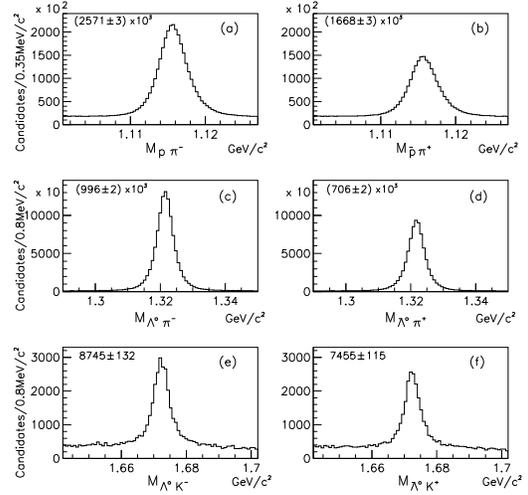,width=7.4cm}}
\caption{$\Lambda^{\circ}$ (upper), $\Xi$ (middle) and $\Omega$ (lower) 
invariant mass plots for the final data samples. Left side figures are particles, right side are antiparticles.}
\label{fig1}
\end{figure}

From fits to a gaussian and a linear background we obtained $2,571,700\pm3,100$ 
$\Lambda^{\circ}$ and $1,669,000\pm2,600$ $\bar{\Lambda}^{\circ}$ from approximately 
$6.5\%$ of the total E791 data sample, $996,200\pm1,900$ $\Xi^-$ and 
$706,600\pm1,700$ $\Xi^+$ and $8,750\pm130$ $\Omega^-$ and $7,469\pm120$ 
$\Omega^+$, these last four being from the total E791 data sample. The final 
data samples for hyperons are shown in Fig.~\ref{fig1}.

\begin{figure}[htb]
\centerline{\psfig{figure=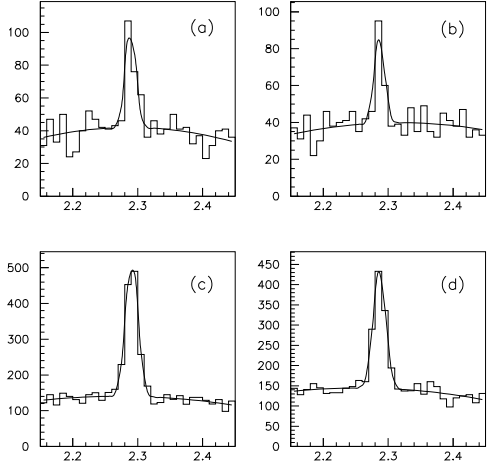,width=7.4cm}}
\caption{$pK^-\pi^+$ and $\bar{p} K^+\pi^-$ mass distributions showing
  clear $\Lambda_c^+$ and $\Lambda_c^-$ signals in each case in $x_F<0$ and $x_F>0$
  regions. From fits with a Gaussian and a parabolic background we 
obtained $122\; \pm 17$ $\Lambda_c^+$ (a) and $92\; \pm 15$ $\Lambda_c^-$ (b) in 
the negative $x_F$ and $903\; \pm 42$ $\Lambda_c^+$ (c) and $702\; \pm 40$ (d) 
$\Lambda_c^-$ in the positive $x_F$ regions.}
\label{fig2}
\end{figure}

For charm baryons, the five targets were used. In most cases, $\Lambda_c$'s 
decayed in air between the target foils, and before entering the silicon 
vertex detectors. All combinations of three tracks consistent with an 
{\it a priori} Cherenkov probability of being identified as a $pK\pi$ and 
c.c. combination were selected for further analysis if the distance between 
$\Lambda_c$ decay vertex to the primary vertex was at least 5 standard 
deviations, and the invariant mass was between $2.15$ and $2.45$ GeV/$c^2$. 
To further enrich the sample, we required the $\Lambda_c$ to decay at least 
five standard deviations downstream of the nearest target foil and 
between one 
and four lifetimes. The $\Lambda_c$ momentum vector, reconstructed from its 
decay products, was required to pass within $3\sigma$ of the primary 
vertex. It was required that  
primary and secondary vertex had acceptable $\chi^2$ per degree of freedom. 
We also required that at least two of the three $\Lambda_c$ decay tracks be 
inconsistent with coming from the primary vertex. The final data sample, 
fitted to a gaussian plus a quadratic background has $1,025\pm45$ $\Lambda_c^+$ 
and $794\pm42$ $\Lambda_c^-$. The invariant mass plot for the $pK\pi$ 
combination is shown in Fig.~\ref{fig2}.


For each baryon - antibaryon pair, an asymmetry both as a function of 
$x_F$ and $p_T^2$ was calculated by means of eq.~\ref{eq1}. Values for 
$N$ ($\overline{N}$) were obtained from fits to the corresponding 
effective mass plots for events selected within specific $x_F$ and $p_T^2$ 
ranges. In all cases, well defined particle signals were evident.

Efficiencies and geometrical acceptances were estimated using a 
sample of Monte Carlo (MC) events produced with the PYTHIA and 
JETSET event generators \cite {pyth}. These events were projected 
through a detailed simulation of the E791 spectrometer and then 
reconstructed with the same algorithms used for data. In the simulation 
of the detector, special care was taken to represent the
behaviour of tracks passing through the deadened region of the drift chambers
near the beam. The behavior of the apparatus and details of the reconstruction
code changed during the data taking and long data processing periods,
respectively. In order to account for these effects, we generated 
the final MC sample into subsets mirroring these behaviors and 
fractional contributions to the final data set. Good agreement between MC
and data samples in a variety of kinematic variables and resolutions was
achieved. We generated $5$ million of $\Lambda^{\circ}$, 
$16.4$ million of $\Xi$, $4.8$ million of $\Omega$, and $7$ million of 
$\Lambda_c$ MC events.

Sources of systematic uncertainties were checked in each case. For 
hyperons we looked for effects coming from changes in the main selection criteria, minimun transverse energy in the calorimeter required in the event trigger, uncertainties in the relative efficiencies for particle and antiparticle, effects of the $2.5\%$

 $K^-$ contamination in the beam, effects of $K^{\circ}$ contamination in the $\Lambda^{\circ}$ sample, stability of the analysis for different regions of the fidutial volume and binning effects. For $\Lambda_c$'s we checked the effect of varying the main

 selection criteria, the effect of the kaon contamination in the beam, 
the contamination of the data sample with $D$ and $D_s$ mesons decaying 
in the $K\pi\pi$ and $KK\pi$ modes and the parametrization of the 
background shape. 

Systematic uncertainties are small and negligible in comparison with 
statistical errors for the $\Lambda_c$ asymmetry. However they are not 
for the hyperons, and are included in the error bars.

\begin{figure}[ht]
\centerline{\psfig{figure=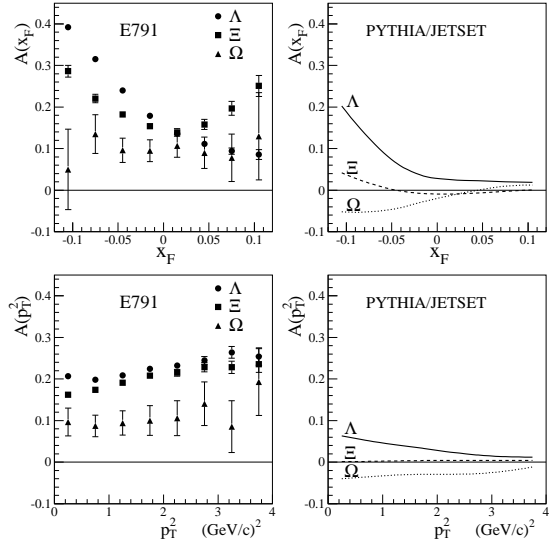,width=7.4cm}}
\caption{$\Lambda^{\circ}$, $\Xi$ and $\Omega$ asymmetries as a function 
of $x_F$ (upper right) and $p_T^2$ (lower right). The asymmetry for 
$x_F$ ($p_T^2$) is integrated over all the $p_T^2$ ($x_F$) range of 
the data set. The left column figures show the predictions of PYTHIA/JETSET.}
\label{fig3}
\end{figure}
\begin{figure}[htb]
\centerline{\psfig{figure=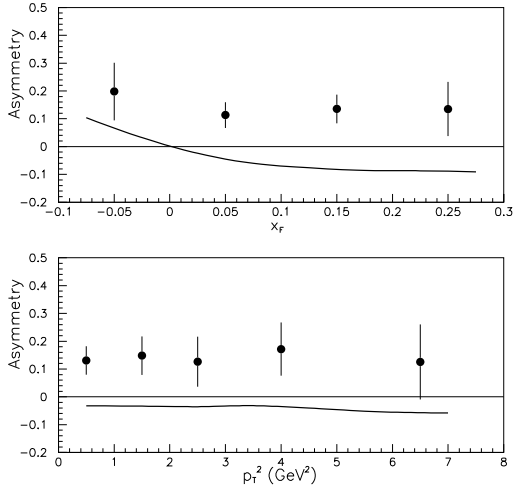,width=7.4cm}}
\caption{$\Lambda_c^+/\Lambda_c^-$ asymmetry as a function of 
$x_F$ (upper) and $p_T^2$ (lower). Full lines are the prediction of 
PYTHIA/JETSET. The asymmetry for $x_F$ ($p_T^2$) is integrated 
over all the $p_T^2$ ($x_F$) range of the data set.
The thin horizontal lines are for reference only. Experimental points are in the center of the corresponding bin.}
\label{fig4}
\end{figure}

Asymmetries in the corresponding $x_F$ ranges integrated over our $p_T^2$ 
and in the corresponding $p_T^2$ range integrated over our $x_F$ range are 
shown in Fig.~\ref{fig3} and \ref{fig4} for hyperons and $\Lambda_c$ baryons 
respectively, in comparison with predictions from the default PYTHIA/JETSET.

We have presented data on hyperon and $\Lambda_c$ production 
asymmetries in the central region for both $x_F>0$ and $x_F<0$. The 
range of $x_F$ covered allowed the first simultaneous study 
of the hyperon and $\Lambda_c$ production asymmetry 
in both the negative and positive $x_F$ regions in a fixed target 
experiment. Our results show, in all cases, a positive asymmetry  
after acceptance corrections over all the kinematical range studied.

Our results are consistent with results obtained by previous 
experiments~\cite{hyperon}. 

Our data shows that leading particle effects play an increasingly important 
role as $\left|x_F\right|$ increases. The non-zero asymmetries measured in 
regions close to $x_F=0$ suggest that energy thresholds for the associated 
production of baryons and mesons play a role in particle antiparticle 
asymmetries.

\begin{figure}[htb] 
\centerline{\psfig{figure=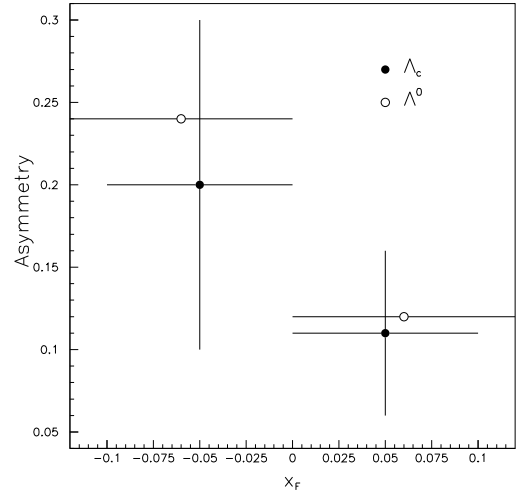,width=7.4cm}}
\caption{Comparison between the $\Lambda^{\circ}$ and $\Lambda_c$ 
asymmetries as a function of $x_F$. The horizontal error bars indicate  the size of the bin in each case.}
\label{fig5}
\end{figure}

On the other hand, the similarity in the $\Lambda^{\circ}$ and $\Lambda_c$ 
asymmetries as a function of $x_F$ (see Fig.~\ref{fig5}) suggest that the 
$ud$ diquark shared between the produced $\Lambda$ baryons and Nucleons in 
the target should play an important role in the measured asymmetry in the 
$x_F<0$ region. However, one expects the $\Lambda_c$ asymmetry to grow 
more slowly than the $\Lambda^{\circ}$ asymmetry due to the mass 
difference between the two particles.

The PYTHIA/JETSET model describes only qualitatively our results, 
which in turn can be better described in terms of a model including 
recombination of valence and sea quarks already present in the initial (colliding) 
hadrons and effects due to the energy thresholds for the associated production of 
baryons and mesons~\cite{silafae}.

\end{document}